\documentclass[twocolumn,showpacs,preprintnumbers,amsmath,amssymb]{revtex4}
\usepackage{dcolumn}

\usepackage{graphicx,amssymb,amsmath}

\begin{document}

\title{Hamiltonian dynamics reveals the existence of
quasi-stationary states for long-range systems in contact with a reservoir}

\author{Fulvio Baldovin and Enzo Orlandini}
\email{baldovin@pd.infn.it, orlandin@pd.infn.it}
\affiliation{
Dipartimento di Fisica and
Sezione INFN, Universit\`a di Padova,\\
\it Via Marzolo 8, I-35131 Padova, Italy
}

\date{\today}

\begin{abstract}
We introduce a Hamiltonian dynamics for the description of long-range
interacting systems in contact with a thermal bath (i.e., in the
canonical ensemble).
The dynamics confirms statistical mechanics equilibrium
predictions for the Hamiltonian Mean Field model and the equilibrium
ensemble equivalence.
We find that long-lasting quasi-stationary states persist
in presence of the interaction with the environment.
Our results indicate that quasi-stationary states are indeed
reproducible in real physical experiments.
\end{abstract}

\maketitle

The statistical mechanics of systems with long-range interactions
is important for a variety of physical applications, including,
e.g., gravitational systems, plasmas, Bose-Einstein condensates
\cite{dauxois}. In such systems, the inter-particle interactions
decay at large distances $r$ as $1/r^\alpha$ with $\alpha\leq d$
(spatial dimension) and ordinary statistical mechanics assumptions
are questioned by nontrivial effects, like persistence of
correlations and non-negligible interface energies. In particular,
the Boltzmann transport equation picture for the approach to
equilibrium is not valid \cite{balescu_1} and long-range
interacting systems may even display inequivalences among
different equilibrium statistical ensembles
\cite{barre_0,barre_1}. Because of these subtleties a privileged
investigation tool is the microscopic Hamiltonian dynamical simulation. 
Hamiltonian dynamics at fixed-energy (microcanonical ensemble) for
a paradigmatic long-range Hamiltonian (see below) directly
connected with  experiments \cite{barre_2} revealed the existence
of long-lived quasi-stationary states (QSS) that finally cross
over to Boltzmann-Gibbs (BG) statistical equilibrium
\cite{latora}. The question arises about the reproducibility of
such QSSs in real physical experiments, where the environment
introduces perturbations that cannot be taken into account in the
microcanonical ensemble. If so, practical advantages could be
obtained by knowing control mechanisms that improve or hinder this
quasi-stationarity. The use of classical numerical prescriptions
(like the Nos\'e-Hoover or the Monte Carlo \cite{frenkel}), to
perform  such an investigation raises some subtle questions, since
these methods implicitly assume the BG equilibrium. Here we
propose a novel, physically transparent, microscopic Hamiltonian dynamics
obtained by coupling the long-range system with a thermal bath
(TB) which simulates the effect of the environment on the system
itself. For sufficiently large time-scales, such dynamics confirms
the equilibrium canonical ensemble predictions.
However, starting with out-of-equilibrium initial conditions, we
discover that long-lived QSSs subsist in presence of the TB,
providing evidence in favor of the reproducibility of QSSs in real
physical experiments. We also discuss the relaxation process
following the QSS and the peculiar behavior of the Boltzmann'
$H$-function.

In a magnetic context, the Hamiltonian Mean Field (HMF) model \cite{konishi}
describes a set of $M$  globally coupled $X Y$-spins
with Hamiltonian
\begin{equation}
H_{HMF}=\sum_{i=1}^M\frac{l_i^2}{2}
+\frac{1}{2M}\sum_{i,j=1}^M\left[1-\cos(\theta_i-\theta_j)\right],
\label{hmf}
\end{equation}
where $\theta_i\in[0,2\pi)$ are the spin angles and $l_i\in\mathbb
R$ their angular momenta (velocities). The presence of the kinetic
term naturally endows the system of spins with an Hamiltonian
dynamics. This Hamiltonian is considered ``paradigmatic'' for
long-range interacting systems \cite{chavanis} since its
equilibrium properties are analytically solvable both in the
microcanonical and in the canonical ensemble
\cite{barre_1,chavanis} and it is representative of the class of
Hamiltonians on a one-dimensional lattice in which the potential
is proportional to
$\sum_{i,j=1}^M\left[1-\cos(\theta_i-\theta_j)\right]/r_{i
j}^\alpha$, where $r_{i j}$ is the lattice separation between
spins and $\alpha<1$ \cite{campa} (notice that the potential in
Eq. (\ref{hmf}) is recovered in the limit $\alpha\to0$). Also,
direct connections with the problem of disk galaxies
\cite{chavanis} and free electron lasers experiments
\cite{barre_2} have been established. Whereas it has been recently
proven that (\ref{hmf}) does not present microcanonical/canonical
inequivalence at equilibrium \cite{barre_1}, its unusual dynamical
features received recently a lot of attention
\cite{latora,chavanis,choi,yamaguchi}. In fact, fixed-energy
dynamical simulations starting with out-of-equilibrium initial
conditions display the existence of longstanding
(infinite-standing in the thermodynamic limit) QSSs appearing
after a ``violent relaxation'' dynamics. During the QSS, phase
functions such as the specific kinetic and potential energies
fluctuate around stationary or quasi-stationary non-equilibrium
average values. In this Letter we introduce a microscopic setup
where the HMF model is in contact with a short-range thermal bath
(TB) in such a way that the thermodynamic limit is achieved with a
negligible interaction energy. Equilibrium dynamics confirms the
microcanonical/canonical equivalence for the HMF model. If the
HMF-TB coupling is weak enough, the relaxation to equilibrium is
still characterized by drastic slowing-downs (QSSs) where also the
system energy fluctuate around quasi-stationary average values.
We discuss in a separate paper \cite{paper_2} the details about the
statistical mechanics of  QSSs in the canonical ensemble,
a question of considerable debate
\cite{latora,chavanis,choi,yamaguchi}.
Here we report that in
Gibbs' $\Gamma$-space their statistical mechanics is obtained
using the classical BG definition of the entropy.

\begin{figure}
\includegraphics[width=0.55\columnwidth]{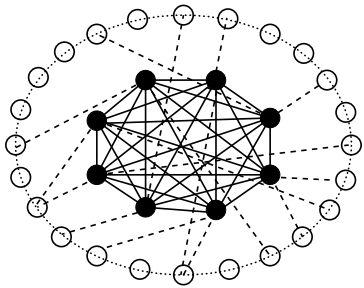}
\caption{ Sketch of the interactions considered in our canonical
setup. Dashed lines mimic the interactions between the HMF- (full
circles) and the TB- (empty circles) spins. Full (dotted) lines
represent the HMF (TB) couplings. } \label{fig_can_setup}
\end{figure}

The TB we consider is characterized by $N\gg M$ equivalent spins
first-neighbors coupled along a chain
\begin{equation}
H_{TB}=\sum_{i=M+1}^{N}\frac{l_i^2}{2}
+\sum_{i=M+1}^{N}\left[1-\cos(\theta_{i+1}-\theta_i)\right],
\label{tb}
\end{equation}
with $\theta_{N+1}\equiv\theta_{M+1}$.
The interaction between (\ref{hmf}) and (\ref{tb}) is
modulated by a coupling constant $\epsilon$:
\begin{equation}
H_{I}=\epsilon\sum_{i=1}^{M}\sum_{s=1}^S\left[1-\cos(\theta_{i}-\theta_{r_s(i)})\right],
\label{interaction}
\end{equation}
where $r_s(i)$ are independent integer random numbers in the interval
$[M+1,N]$. In this way, each HMF-spin is in contact with a set of $S$ different
TB-spins chosen randomly along the chain (see
Fig. \ref{fig_can_setup}). This set is specified as initial condition
and remains fixed during the dynamics. The total Hamiltonian
$H=H_{HMF}+H_{TB}+H_I$ defines then a microcanonical system
(constant energy $E$), in which the energy of the HMF model can
fluctuate. In our approach, the temperature is defined by (twice)
the specific kinetic energy and we expect the TB to maintain a
constant temperature about which the HMF model thermally
equilibrates. By assuming a ``surface-like effect'' $S\sim
M^{\gamma-1}$ (with $0<\gamma<1$), we make sure that the
interaction energy, $E_I\sim M^\gamma$, satisfy
$E_{HMF}\sim M$ ($E_{TB}\sim N\gg M$), thus ensuring a well
defined thermodynamic limit. For the present results we chose
$N=M^2$ and $S=10^5M^{-1/2}$. To integrate the equation of motion
we use a velocity Verlet \cite{frenkel} algorithm with an
integration step guaranteeing conservation of total energy within
an uncertainty of $\Delta E/E \simeq 10^{-5}$.

\begin{figure}
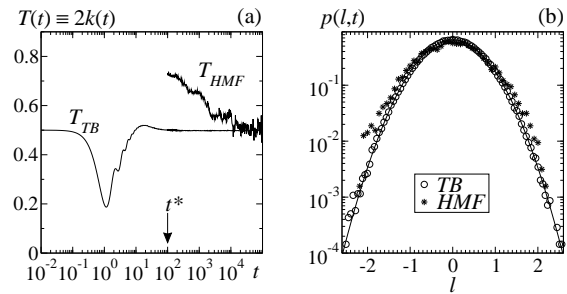
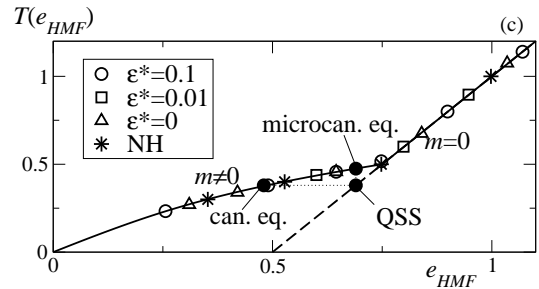

\includegraphics[width=0.45\columnwidth]{temp_bath.eps}
\includegraphics[width=0.45\columnwidth]{pdf_vel_eq.eps}\\
\vspace{0.25cm}
\includegraphics[width=0.90\columnwidth]{caloric.eps}
\caption{
  (a): Time evolution of the $T_{HMF}$ and $T_{TB}$
  temperatures for $M=10^3$, $\epsilon=0.01$ and $T_0=0.5$.
  Initially, 
  $p_{HMF}(l,0)=\exp(-l^2/2T(t^*))/\sqrt{2\pi T(t^*)}$ with $T(t^*)=0.7$.
  (b): Velocity PDF at $t\gg t^*$.
  Solid line is $p_{TB}(l,0)$.
  (c): Caloric curve. Solid line is the BG equilibrium
  and dashed line is the prolongation of the ordered
  phase to subcritical energies.
  Empty symbols are the average value of
  $e_{HMF}(t)$ at equilibrium.
  Stars refer to Nos\'e-Hoover
  simulations. Full circles correspond to the QSS studied in the paper
  and to the  microcanonical and canonical equilibrium 
  obtained as $t\to\infty$.
}
\label{fig_eq}
\end{figure}

Let \mbox{$e_{HMF}\equiv E_{HMF}/M=[k_{HMF}+(1-m_{HMF}^2)/2]$},
where  $k_{HMF}\equiv\sum_{i=1}^Ml_i^2/2M$ and $m_{HMF}\equiv
|\sum_{i=1}^M(\cos\theta_i,\sin\theta_i)|/M$ are respectively the
specific kinetic energy and the magnetization of the system. It is
known that at $e_{HMF}=0.75$ and temperature $T_{HMF}=0.5$ (in
natural dimensionless units) a continuous phase transition occurs
separating a disordered ($m_{HMF}=0$) phase from a ferromagnetic
one \cite{konishi,chavanis}. Here we show that our Hamiltonian
dynamics confirms such equilibrium predictions. The width $T_0$ of
the Maxwellian PDF for the initial TB-velocities
$p_{TB}(l,0)=\exp(-l^2/2T_0)/\sqrt{2\pi T_0}$ is a control
parameter through which we set the TB temperature. In fact, after
a transient relaxation ($0\leq t<t^*\sim100$) the TB reaches its
own equilibrium at the target temperature $T_0$ (i.e.
$2k_{TB}(t)\simeq T_0\;\forall t>t^*$). At $t=t^*$ we then switch
on the HMF-TB coupling, $H_I$, by setting
$\epsilon(t)=\epsilon^*\geq0\;\forall t\geq t^*$. For
$\epsilon^*=0$, $H_I=0$, and the scheme reproduces the
microcanonical dynamics of the HMF. The setup was tested for many
different initial conditions of the HMF model with $10^2\leq M\leq
10^4$ and $0.005\leq\epsilon^*\leq0.1$. In all cases, for $t\gg
t^*$, the system reaches the thermal equilibrium characterized by
$2k_{HMF}(t)\simeq T_0$ (Fig \ref{fig_eq}a) a velocity PDF
$p_{HMF}(l,t)\simeq p_{TB}(l,0)$ (Fig. \ref{fig_eq}b), and an
equilibrium magnetization. The relaxation to equilibrium could
last very long and typically occurs through a number of drastic
slowing-downs during which the average value of $T_{HMF}$ is
constant or almost constant (plateaux in Fig. \ref{fig_eq}a).
Similar effects were found in \cite{morita} using a stochastic
dynamics (see also \cite{chavanis} for a stochastic canonical version
of  the HMF, named Brownian mean field). 
By varying $T_0$, we obtain an estimate of the caloric
curve in excellent agreement both with the BG equilibrium
prediction and with the microcanonical simulation, even at the
critical temperature and independently of $\epsilon^*$ (Fig.
\ref{fig_eq}c). These findings confirm the equilibrium ensemble
equivalence \cite{barre_1} solely on the basis of Hamiltonian
dynamics. For close to equilibrium initial conditions, the 
Nos\'e-Hoover dynamics \cite{frenkel} reproduces the caloric curve
(Fig. \ref{fig_eq}c) as well.

\begin{figure}
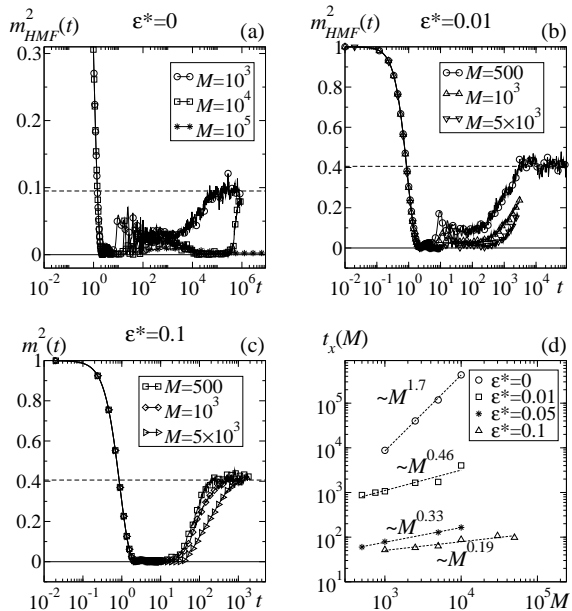

\includegraphics[width=0.45\columnwidth]{mag_eps0.eps}
\includegraphics[width=0.45\columnwidth]{mag_eps0p01.eps}\\
\vspace{0.25cm}
\includegraphics[width=0.45\columnwidth]{mag_eps0p1.eps}
\includegraphics[width=0.45\columnwidth]{t_vs_M.eps}
\caption{
  Microcanonical (a) and canonical (b,c) QSSs in terms of
  $m_{HMF}^2$. The curves have been obtained by averaging
  over a number $I$ of initial conditions following the probability
  distributions described in the text. $I$ varies from a maximum of $20$
  for small $M$ down to $5$ for big $M$.
  (d): Log-log plot of $t_x$ as a function of the system size $M$.
  The QSS life-time $t_x$ has been defined as the time at which a
  phase function ($T_{HMF}$ for $\epsilon^*=0$ and
  $e_{HMF}$ for $\epsilon^*\neq0$) changes $10\%$ of its stationary average
  value.
}
\label{fig_qss_1}
\end{figure}

We now turn to the non-equilibrium properties of the model by
setting for the $HMF$ system out-of-equilibrium initial
conditions. In particular,  at $t=t^*$, we consider a delta
distribution for the angles ($p_{HMF}(\theta,t^*)=\delta(0)$ so
that $m_{HMF}^2(t^*)=1$) and a uniform distribution for the
velocities, $p_{HMF}(l,t^*)=1/2\bar l,\;l\in[-\bar l,\bar l]$,
with $\bar l\simeq2.03$ ($e_{HMF}(t^*)\simeq0.69$). If
$\epsilon^*=0$, we verified the known result that for such initial
conditions the system, after a fast process, is dynamically
trapped into a QSS \cite{chavanis,latora,choi,yamaguchi} (Fig.
\ref{fig_qss_1}a). The initial ($t\lesssim t^*+1$) violent
relaxation corresponds to a quick mixing of the spins in the
single-particle $\mu$-space \cite{chavanis}. The QSS is then
characterized by $m_{HMF}^2\simeq0$ ($T_{HMF}=0.38$) for
$M\to\infty$ (zero force) and a lifetime $t_x$ that increases as a
power of $M$ \cite{latora,yamaguchi}. With respect to such QSSs, a
crucial issue is to see whether they survive when the coupling
between the HMF and the TB is switched on \cite{choi}. To address
this point, we consider $\epsilon^*\neq0$ but keeping the
non-equilibrium initial conditions described above. The
TB-temperature is first fixed at $T_0=0.38$. The time dependence
of $m^2_{HMF}$ (Fig. \ref{fig_qss_1}b,c) suggests that the QSSs
indeed exist even in the canonical setup and independently of
$\epsilon^*$  (if $\epsilon^*$ is small enough). Denoting by $t_x$
the life time of the QSSs we found that $t_x\sim M^\eta$, with
$\eta$ that tends to zero as $\epsilon^*$ increases and to the
microcanonical estimate given in \cite{yamaguchi} as
$\epsilon^*\to 0$ (Fig. \ref{fig_qss_1}d). Preliminary evidences
\cite{preparation}, suggest that $t_x$ is also influenced by  the
``surface effect'' parameter $\gamma$. We remark that during the
QSS the long-range system does not thermalize with the TB. For
example, a consistent change ($10\%$) of $T_0$ does not alter
$T_{HMF}$ and even the subset of TB-spins
in direct
contact with the HMF model remains at $T_0$ \cite{paper_2}.
However, energy fluctuations are significantly larger than those
due to the algorithm precision ($\Delta E_{HMF}/E_{HMF}\simeq
4\times10^{-2}$ for $M=10^3$) \cite{paper_2}. This distinguishes
the canonical QSSs from the microcanonical ones. During these
QSSs, the HMF model are in a partial equilibrium state at a
temperature (specific kinetic energy) which is not the one of the
TB \cite{paper_2}. Perhaps this is one reason why a Nos\'e-Hoover
dynamics with the same out-from-equilibrium initial conditions is
not capable to reproduce the relaxation to equilibrium. In fact,
we verified \cite{preparation} that in such a case a Nos\'e-Hoover
dynamics displays very strong fluctuations of the dynamical variables (e.g.,
$E_{HMF}$) that do not decay with time. Another
important remark is that classical assumptions in mesoscopic
stochastic equations seem to rule out the existence of such
canonical QSS. In fact, a stability analysis applied to a
Fokker-Planck description of the HMF model in both ensembles
(canonical and microcanonical) shows that anomalous velocity PDFs
are (neutrally) stable only in the microcanonical ensemble
\cite{choi}.

\begin{figure}
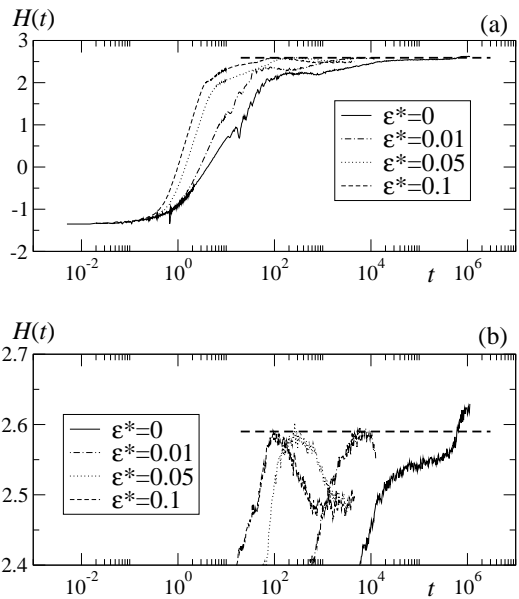

\includegraphics[width=0.90\columnwidth]{eta_a.eps}\\
\vspace{0.25cm}
\includegraphics[width=0.90\columnwidth]{eta_b.eps}
\caption{Time dependence of the $H$-function in the microcanonical
  ($\epsilon^*=0$) and canonical ($\epsilon^*\neq0$) ensemble for
  $M=10^4$ and $T_0=0.38$. The dashed line indicate the common maximum
  of $H$ for $\epsilon^*\neq0$ reached during the QSS.
}
\label{fig_eta}
\end{figure}

The occurrence of canonical QSSs points towards
an extension of the ensemble equivalence to some
aspects of the non equilibrium properties.
We find on the other hand that there is a substantial
microcanonical/canonical {\it inequivalence} in the relaxation to
equilibrium process that follows the QSS.
For example, the final equilibrium
specific magnetization changes by a factor $4$ going from the
microcanonical (Fig. \ref{fig_qss_1}a) to the canonical (Figs.
\ref{fig_qss_1}b,c) simulations, independently of $\epsilon^*$.
A further indication of this
inequivalence is given by the time evolution of the $H$-function,
\mbox{$H(t)\equiv-\int_{-\infty}^{+\infty}d
l\int_{0}^{2\pi}d\theta\; p(l,\theta,t)\ln(p(l,\theta,t))$},
in the two ensembles. Indeed, if  during the microcanonical
dynamics $H(t)$ on the average increases, reaching its maximum at
equilibrium, in the $\epsilon\neq0$ canonical ones it displays a
maximum during the QSS and then {\it decreases} towards the
$T_0=0.38$ equilibrium value (See Fig. \ref{fig_eta}). These
unconventional behaviors with respect to the $H$-theorem are
consequences of the different relaxation dynamics in the two
ensembles, i.e. fixed-energy and fixed-temperature (full circles
in Fig. \ref{fig_eq}c).

In summary, we introduced a Hamiltonian canonical setup for
long-range interacting systems through a coupling with a
short-range interacting TB. The coupling is described by a
parameter that can be tuned continuously to provide a unified
description of microcanonical and canonical ensembles. By applying
our scheme to the HMF model we verified its capability to
reproduce the equilibrium BG statistics in both ensemble (ensemble
equilibrium equivalence). The major feature of this setup is that
it is based solely on microscopic Hamiltonian dynamics. This is a novelty with
respect to previous approaches, that allows an unbiased dynamical
description of the non equilibrium properties of such system. As a
result we found that, if the coupling with the TB is weak enough
and we start from out-of-equilibrium initial conditions, the
dynamics reveals the existence of quasi stationary states in the
canonical ensemble. These QSSs are  reminiscent of the
microcanonical ones \cite{chavanis,latora,choi,yamaguchi} in the
sense that, for example, their lifetime diverges with the system
size $M$ in a power law fashion. On the other hand in presence of
the TB, the life-time of the QSSs is influenced by the parameters
controlling the interaction between long-range system and TB. This
could be useful for an experimentalist who is willing to enhance
or hinder the quasi-stationary behavior \cite{barre_2} and could
also be of some importance in the understanding of the dynamical
evolution of quasi-stationary structures, e.g., in galaxies
\cite{chavanis}, or in other long-range interacting systems. The
presence of canonical QSS extends the notion of ensemble
equivalence from equilibrium to some non equilibrium properties. A
substantial microcanonical/canonical {\it inequivalence} is found
in the relaxation to equilibrium process following the QSS and it
is clearly revealed by a dramatic change in the time dependence of
the Boltzmann $H$-function. Of course, a more detailed statistical
description (and interpretation) of the canonical QSS is needed
and we are confident that our unbiased set up will be a useful
tool for this achievement\cite{paper_2} not only with respect to
the HMF model, but also to other Hamiltonian long-range systems
exhibiting either dynamical peculiar features \cite{morita2} or
equilibrium ensemble inequivalence \cite{barre_0}.

\section*{Acknowledgments}
We thank A.L. Stella and S. Ruffo for useful remarks.
FB also acknowledges collaboration with C. Tsallis and
L.G. Moyano
with whom initial interest for this problem has been
shared,
and F. Leyvraz, A. Rapisarda and H. Touchette for discussions.

\end{document}